\documentclass[wcp]{jmlr}

\usepackage{longtable}

\usepackage{booktabs}

\usepackage{lineno}

\usepackage{multirow}
\usepackage{subcaption}

\makeatletter
\let\Ginclude@graphics\@org@Ginclude@graphics 
\makeatother

\jmlryear{2026}
\jmlrworkshop{ACML 2026}

\title[DPonWVSAlign]{Differential Privacy Protected Preference Alignment on Survey Derived Data}

\author{\Name{Xiuyuan Qi} \Email{qixy2025@shanghaitech.edu.cn}\\
    \Name{Yi Wang} \Email{wangyi2024@shanghaitech.edu.cn}\\
    \Name{Wenjie Wang}\thanks{Corresponding author.} \Email{wangwj1@shanghaitech.edu.cn}\\
    \addr ShanghaiTech University}

\begin{document}

\makeatletter
\let \@jmlrpages \@empty
\makeatother

\maketitle

\begin{abstract}
Preference alignment enables language models to better reflect human values, but the preference data contains sensitive information, especially when derived from surveys about political beliefs, social norms, religion, gender attitudes, and cultural values. This paper studies privacy-preserving alignment on questionnaire-derived data from the World Values Survey, where supervision takes the form of ordinal labels rather than standard pairwise rankings. Existing efforts mainly focus on general privacy protections upon model training, while data-specific methods for survey data remain unexplored. We first show the vulnerability of aligned models, then formulate the task as rating prediction and introduce label differential privacy. Building on this formulation, we construct a differential private pipeline utilizing three privacy mechanisms: randomized response, exponential mechanism, and DP-SGD. Experiments show that simple label-level mechanisms outperform gradient-level methods, and proper aggregation function for preprocessing also make large contribution. The new fine-grained private comparison motivates explorations toward language model alignment with formal privacy guarantees.
\end{abstract}

\section{Introduction}
Model alignment is an important technique for language models (LMs) to generate high quality responses that meet delicate preferences of users. Aligned models are expected to generate responses that better reflect human values and to improve helpfulness and harmlessness in downstream tasks. Most post-training methods rely on human-labeled preference data consisting of a query, multiple candidate answers, and human-provided ranking signals~\cite{rafailov2024directpreferenceoptimizationlanguage, DBLP:journals/corr/SchulmanWDRK17, shao2024deepseekmathpushinglimitsmathematical}. However, such queries are often subjective, and annotators inevitably rely on their own knowledge, values, and opinions when making judgments. As a result, preference data may contain sensitive information, including individual value judgments and behavioral tendencies. This makes privacy protection an important requirement in model alignment for service providers.

The World Values Survey (WVS)~\cite{haerpfer2022wvs} is a global research project that investigates people’s values and beliefs across different cultures and societies. It provides high quality data of preferences of individuals from more than 120 societies. Although WVS respondents are anonymized, privacy risks may still arise when survey derived preference signals are used to train language models.  Research on group privacy~\cite{Floridi2017} and linkage attack~\cite{DBLP:journals/popets/PowarB23} indicate that attackers could still exploit the aligned model for data source inference, confidential attribution prediction, or discover sensitive information including political preferences, ethical values, social biases, etc.

\begin{figure}
\centering
\begin{minipage}[h]{0.8\textwidth}
\begin{center}
\parbox[b]{0.4\linewidth}{
    \centering
    \includegraphics[width=\linewidth]{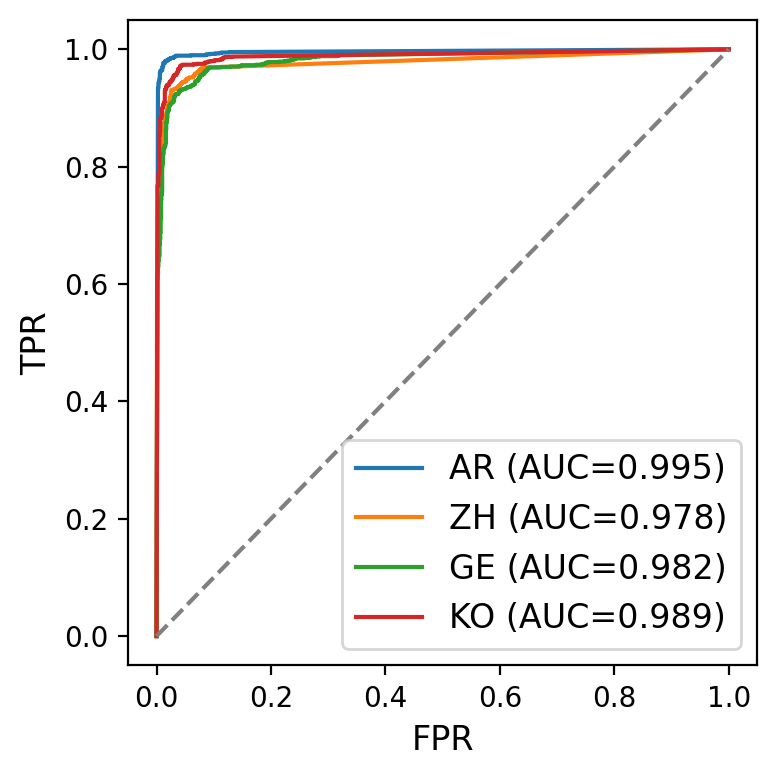}
    \small (a) kNN
}
\hfill
\parbox[b]{0.4\linewidth}{
    \centering
    \includegraphics[width=\linewidth]{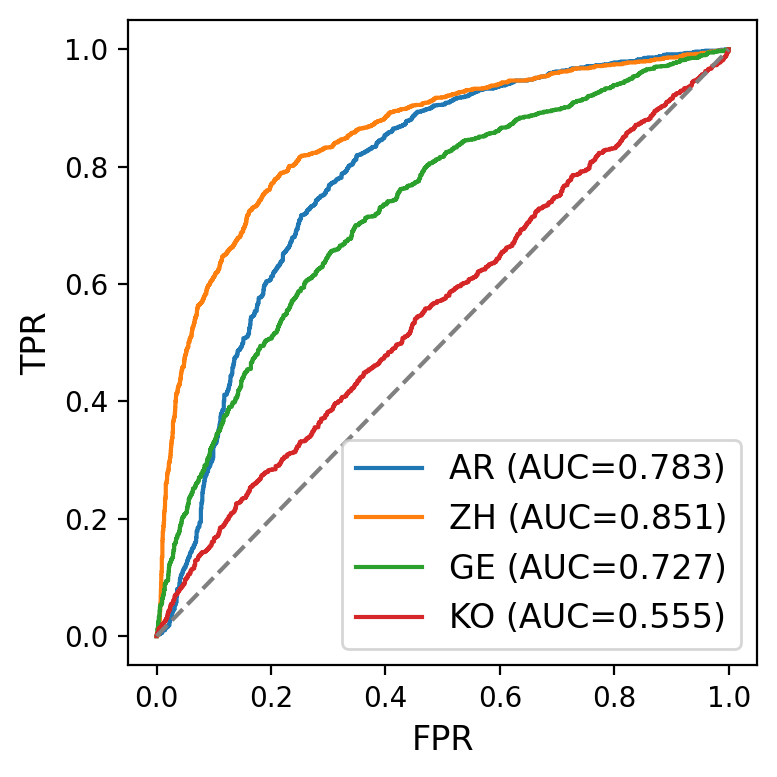}
    \small (b) Logistic Regression
}
\end{center}
\end{minipage}
\caption{Receiver Operating Characteristic (ROC) curve of MIA attack with different methods on English aligned model.}\label{fig:roc}
\end{figure}

To illustrate this privacy risk, we conducted an empirical study: a language model is fine-tuned with averaged answers of different regions from WVS data. An attacker wants to reconstruct the preference vectors and determine from which region a specific record \(r\) comes. The attack simulates the group-level preference-distribution inference~\cite{Floridi2017}. We evaluated the risk of inference with ROC curves in Fig.~\ref{fig:roc}, illustrating high probability of an attacker distinguishing the source region of any in-distribution record \(r\).

Furthermore, differential privacy~\cite{Dwork2006CalibratingNT} provides a formal privacy framework with mathematical guarantees and has become a common standard for privacy protection. Various DP methods have been proposed for different data structures and machine learning tasks. However, WVS contains ordinal preference labels whose structure is inconsistent with traditional pairwise alignment data. This motivates us to evaluate the effectiveness and efficiency of existing DP mechanisms in questionnaire-based preference alignment and to develop a protection strategy that achieves strong privacy guarantees with minimal utility degradation.

In this work, we adapt three classical DP mechanisms to the alignment task, illustrate the privacy-preserving workflow, and evaluate their privacy-utility performance. Since sensitive information in our setting is primarily contained in preference labels rather than query text, we exploit label-level DP mechanisms~\cite{2021labelDP} tailored to questionnaire-derived supervision. Our contributions are summarized as follows:
\begin{enumerate}
    \item We formulate a privacy-preserving preference alignment with questionnaire-based group preferences.
    \item We develop a simple and flexible DP protection pipeline that adapts standard alignment mechanisms to aggregated preference supervision and provides formal privacy guarantees.
    \item We empirically compare DP-SGD, randomized response, and the exponential mechanism on survey derived data alignment task, showing that lightweight label-level mechanisms achieve better privacy-utility trade-offs than the record-level method.
\end{enumerate}

\section{Related works}
\subsection{Preference Alignment}
Preference alignment aims to steer language models toward outputs that better match human intentions, preferences, and values. In practice, alignment is commonly used to improve instruction following, helpfulness, harmlessness, honesty, and safety. Mainstream alignment pipelines usually begin with supervised fine-tuning on instruction-response demonstrations, and then further optimize models with preference feedback, e.g., reward modeling with reinforcement learning from human feedback (RLHF)~\cite{humanPreferenceLearningSurvey}, DPO~\cite{rafailov2024directpreferenceoptimizationlanguage}, and GRPO~\cite{shao2024deepseekmathpushinglimitsmathematical}.

A key challenge is that human preferences are not universal. Existing LLMs are largely aligned to Western values and lack sufficient cultural divergence under different language and cultural settings~\cite{wang-etal-2024-countries}. This limitation is especially problematic for value-laden tasks, where acceptable answers may vary across societies. Recent work like CultureLLM~\cite{Li2024CultureLLM} studies culture-aware alignment, which aims to adapt model behavior to different cultural groups rather than assuming a single global preference function. We investigated in a similar scenario raised in CultureLLM which leverages a cross-cultural survey resources: the World Values Survey (WVS). Compared with standard PPO or DPO datasets, however, WVS-style data has a different structure: responses are ordinal survey labels from population-level response distributions rather than pairwise rankings.

More importantly, existing culture-aware alignment work generally focuses on alignment utility and cultural consistency, but few provides formal privacy guarantees. This motivates our study of privacy-preserving preference alignment with questionnaire-style cultural data.

\subsection{Differential Privacy}
Differential privacy provides a formal framework for mitigating privacy risks with quantifiable guarantees and robustness to arbitrary auxiliary information~\cite{Dwork2006CalibratingNT}. In machine learning, a representative DP approach is DP-SGD~\cite{Abadi2016DeepLW}, which is widely used for private model training and fine-tuning. However, this protection is applied to the entire training record and optimization process, which can be overly conservative and computationally expensive when only part of the data is sensitive.
Other DP variants provide more targeted protection for specific data structures. Selective DP and data-centric approaches further aim to protect only privacy-sensitive tokens, attributes, or variables in structured inputs~\cite{shi-etal-2022-selective}. Label differential privacy~\cite{2021labelDP} protects sensitive labels while leaving input features unchanged, and is suitable when labels encode private information but inputs are public or less sensitive. Although these approaches reduce unnecessary perturbation, they are not directly designed for questionnaire-based preference alignment. Consequently, directly applying existing DP mechanisms leaves open the question of how to achieve a favorable privacy-utility trade-off.

\subsection{WVS}
The World Values Survey (WVS)~\cite{haerpfer2022wvs} is a global research project that investigates people’s values and beliefs across different cultures and societies. It has been conducted in seven waves over the past several decades, covering more than 120 societies and representing over 94.5\% of the world’s population. WVS collects responses on a broad range of topics related to values, beliefs, and social norms. Such large-scale and culturally diverse data provides strong potential for aligning LMs with human values. Existing work~\cite{Li2024CultureLLM} using WVS demonstrated notable improvement on downstream alignment tasks including identification of hate speech, bias and offensive language.

The survey consists of 290 questionnaire items, each associated with a response scale ranging from ``strongly disagree'' to ``completely agree'', rather than the query-answer-ranking format commonly used in preference alignment and therefore do not share the same alignment pipeline. More importantly, the highly sensitive information makes privacy protection curtail when utilizing such data, which is not discussed in existing works.

\section{Problem Statement}\label{sec:problem}
Given alignment data derived from questionnaire responses, our objective is to learn an aligned model \(\theta\) that maximizes alignment utility while satisfying DP constraint.

\subsection{WVS-based Rating Prediction Task}
The problem of using questionnaire-style data for privacy-preserving preference alignment on LMs differs from conventional alignment settings, and we follow a pipeline raised in~\cite{Li2024CultureLLM}. Specifically, the World Values Survey data can be formulated as follows. \( Q = [q_i]_{i=1}^{n} \) denote a set of \(n\) value-related questions, and let \( \mathcal{R} =[r_{ij}]\in\mathbb{R}^{n\times m}\)denote the response matrix with \(n\) questions and \(m\) participants. Each entry \(r_{ij}\) belongs to a discrete response space \(S_i\) (e.g., a scale from 1 to 10), corresponding to survey-scale answers such as levels of agreement. 

The training dataset \(\mathcal{D}= \{(x_i, y_i)\}_{i=1}^{N} \) includes \(N=n\cdot a\) records, where \(x\) is constructed by augmenting each \(q\) for \(a\) times, and applying a query template. A sample of \(x\) is shown below.
\begin{quote}
Give me the answer from 1 to 4: In terms of your belief, how recurrent is the occurrence of journalists delivering fair election coverage in this country's elections? 1. Very often 2. Fairly often 3. Not often 4. Not at all often. You can only choose one option. 
\end{quote}
The corresponding label \(y\) represents aggregated response, formally:
\[
y_i=aggr(r_{i1},...,r_{im})
\]
where \(aggr\) could be mean, median or argmax.
We formulate value alignment on such questionnaire data as a multi-classification task: for each query \(x_i\), the model outputs a probability distribution over the discrete response space \(S\), and predicts the final label by
\[
\hat{y}_i = \arg\max_{y \in S} p_{\theta}(y \mid x_i),
\]
where \(p_{\theta}(y \mid x_i)\) denotes the predicted probability of response \(y\) given input \(x_i\). To constrain the output space to the discrete response set \(S\), we replace the LM head of the language model with a classification head. This can serve as a proxy for the value tendencies encoded in the language model, as the final hidden representation used for prediction captures global semantic and attitudinal information.
The model is optimized with a standard cross-entropy loss:
\[
\mathcal{L}(\theta) = - \sum_{i=1}^{N} \log p_{\theta}(y_i \mid x_i).
\]
This formulation provides a unified prediction objective for both the non-private and privacy-preserving settings. Different privacy mechanisms do not change the task itself, but instead modify either the supervision signal or the optimization process, which will be introduced in the next subsection.

\subsection{Threat Model}
\begin{figure}
\begin{center}
    \includegraphics[width=0.7\textwidth]{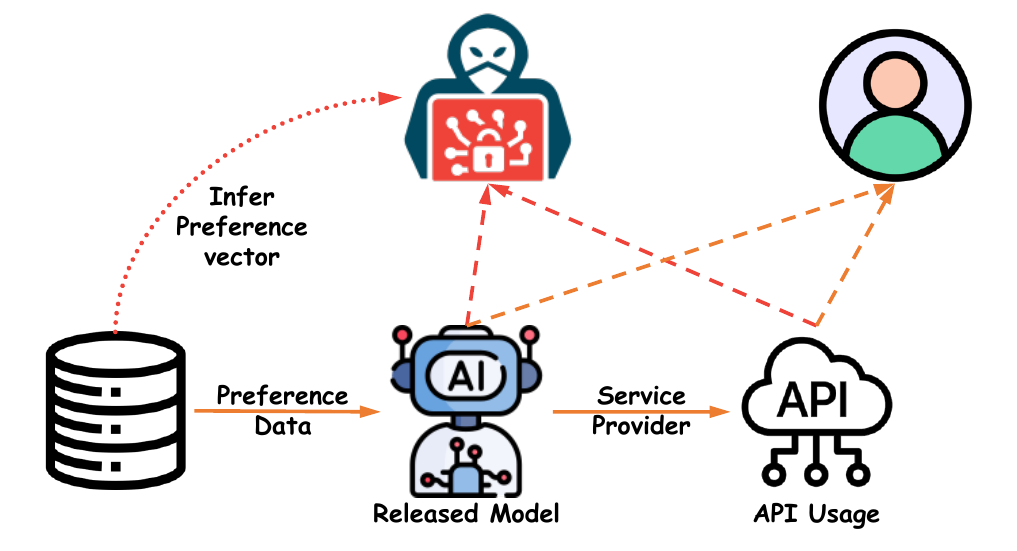}
    \caption{Threat model of privacy attacks in aggregated preference alignment scenario.}
    \label{fig:threatmodel}
\end{center}
\end{figure}

According to the taxonomy described in~\cite{Rigaki2023PrivacyAttacks}, we describe our threat model as a white-box attack against a centralized training paradigm. The attacker has access to target model parameters after alignment, but training data samples and train-time internal state like gradient are kept secret. Fig.~\ref{fig:threatmodel} illustrates the threat model.
\subsection{Privacy Constraint}
In our setting, let \(\mathcal{D}=\{(x_i,y_i)\}_{i=1}^{N}\) denote the training dataset, we define two neighboring datasets as follows:  \(\mathcal{D}\) and \(\mathcal{D}'\) are neighboring if there exists an index \(j\) such that
\[
\left\{\begin{array}{cc}
     y_i=y_i'&\text{if}\; i\ne j\\
     y_i\ne y_i'&\text{if}\;i=j
\end{array}
\right.
\]
and \(x_i=x_i'\;\forall i\in[1,N]\), since all \(x_i\) are derived from original WVS question \(q_i\). Unlike standard record-level DP, which protects the presence or absence of an entire training example, we adopt label DP which protect privacy information of preference label \(y\).

\paragraph{\((\epsilon,\delta)\)-Label Differential Privacy.~\cite{2021labelDP}}
A randomized alignment algorithm \(\mathcal{M}\), whose output space consists of all possible aligned models, satisfies \((\epsilon,\delta)\)-label differential privacy if, for any pair of label-neighboring datasets \(D\) and \(D'\), and for any measurable set of outputs \(\mathcal{O}\), we have
\[
\Pr[\mathcal{M}(\mathcal{D})\in \mathcal{O}]
\leq
e^{\epsilon}
\Pr[\mathcal{M}(\mathcal{D}')\in \mathcal{O}]
+
\delta .
\]
This definition ensures that replacing one ordinal preference label with another valid label in the response space has only a bounded influence on the distribution of the final aligned model. When \(\delta=0\), the above definition reduces to pure \(\epsilon\)-label differential privacy:
\[
\Pr[\mathcal{M}(\mathcal{D})\in \mathcal{O}]
\leq
e^{\epsilon}
\Pr[\mathcal{M}(\mathcal{D}')\in \mathcal{O}] .
\]
Label DP is qualified for  our questionnaire-based alignment setting, as the sensitive object is the ordinal preference label rather than the query text. By leaving the query unchanged, it avoids excessive computational cost and overly conservative privacy constraint.
In  addition, the label-level DP constraint can extend to respondent-level privacy consisting of \(n\) records, with an \((n\cdot\epsilon,0)\) additional privacy budget for RR and EM~\cite{}, and larger cost for DP-SGD~\cite{}. Since respondent identity is removed from each record, the practical gap between the two granularities is constrained, and we leave the study to future works.  

\section{Methodology}\label{sec:method}

\subsection{Privacy-Preserving Alignment Pipeline}
We study privacy-preserving preference alignment under the WVS-based rating prediction task defined in Section~\ref{sec:problem}, and our goal is to incorporate label-level privacy protection into the alignment pipeline. 

The overall workflow consists of four stages:
\begin{itemize}
    \item \textbf{Raw response collection.} We start from the original WVS responses \(\mathcal{R}\), where each questionnaire item \(q\) is associated with a set of ordinal answers \(d\) from participants in a target group. These raw responses contain the group preference information to be protected.

    \item \textbf{Label construction.} For each questionnaire item, the raw responses are converted into a single supervision label \(y\) through an aggregation or private sampling procedure. In the non-private pipeline, this step uses a deterministic aggregation function such as mean, median, or argmax.

    \item \textbf{Training data formatting.} Each questionnaire item is rewritten and inserted into a fixed prompt template to form the model input \(x\), and the constructed label is used as the prediction target. This produces a query-label training dataset \(\mathcal{D}\) for the rating prediction task.

    \item \textbf{Model fine-tuning.} A language model with a classification head is fine-tuned on the constructed training dataset.
\end{itemize}

\begin{figure}[htbp]
\begin{center}
    \includegraphics[width=0.9\textwidth]{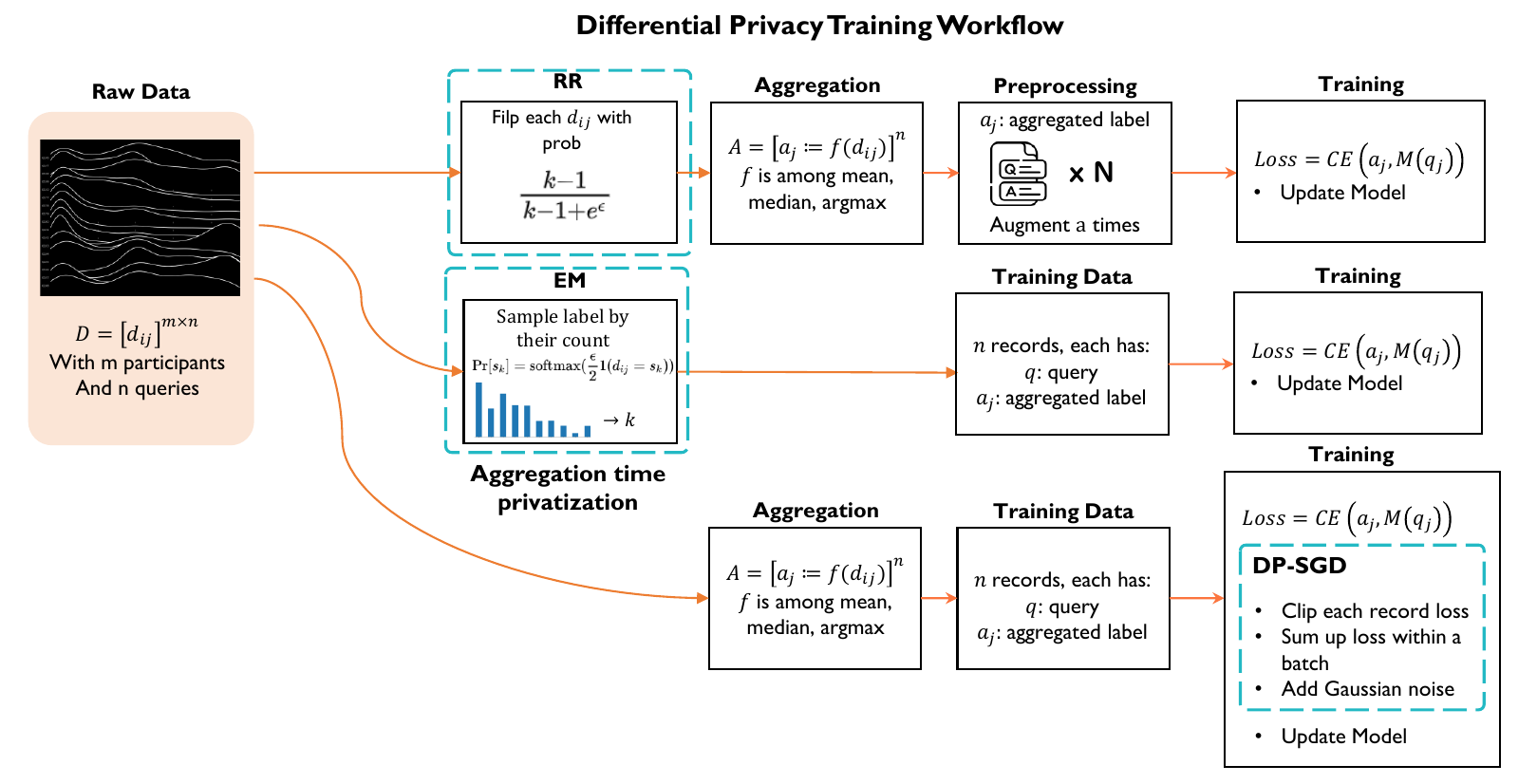}
    \caption{Overall workflow of the three privacy-preserving alignment methods. The shared modules are query construction, training data formation, and model training, while the dashed regions indicate the stages where each privacy mechanism is applied.}
    \label{fig:dp_workflow}
\end{center}
\end{figure}

We consider three privacy mechanisms applied at different stages of this pipeline: randomized response (RR), exponential mechanism (EM), and DP-SGD. As illustrated in Fig.~\ref{fig:dp_workflow}, RR and EM protect the supervision signal before model training phase, while DP-SGD protects the optimization process during fine-tuning. In addition, RR and DP-SGD is combined with three different aggregate functions, leading to up to 7 settings under evaluation. This allows us to make a comprehensive comparison of label-level privacy protections under the same alignment task.
\paragraph{Randomized Response.~\cite{Warner1965RandomizedRA}}
Randomized response applies label-level perturbation after the deterministic aggregation step. For each record \((x_i,y_i)\), RR replaces the label \(y_i\) with a privatized label \(\tilde{y}_i\). Let \(K_i = |S_i|\) denote the number of possible labels. The privatized label is sampled as
\[
\left\{
\begin{array}{cc}
    \Pr(\tilde{y}_i = y_i)
    =
    \frac{e^{\epsilon}}{e^{\epsilon}+K_i-1}\\
    \Pr(\tilde{y}_i = k)
    =
    \frac{1}{e^{\epsilon}+K_i-1}
\end{array}
\right.
\]
for any alternative label \(k \neq y_i\). The resulting private training dataset is
\[
\tilde{\mathcal{D}}_{\mathrm{RR}}
=
\{(x_i,\tilde{y}_i)\}_{i=1}^{N}
\]
RR is simple and directly satisfies label-level privacy, but it treats all incorrect labels uniformly and does not explicitly exploit the ordinal structure of the response space.

\paragraph{Exponential Mechanism.~\cite{McSherry2007MechanismDV}}
The exponential mechanism protects the label construction stage by replacing deterministic aggregation with a differential private sampling rule. For each candidate label \(k \in S_i\), we define a utility function based on its empirical support in the response vector:
\[
u_i(k)=\frac{1}{n_i}\sum_{j=1}^{n_i}\mathbb{I}(r_{ij}=k)
\]
This utility measures how strongly label \(k\) is supported by the group responses. Since changing one individual response can change this utility by at most \(1/n_i\), its sensitivity is
\[
\Delta u_i = \frac{1}{n_i}.
\]
The private label is sampled according to
\[
\Pr(\tilde{y}_i = k)
\propto
\exp\left(
\frac{\epsilon u_i(k)}{2\Delta u_i}
\right),
\qquad k \in S_i.
\]
Thus, labels with higher empirical support are more likely to be selected, while the sampling procedure satisfies differential privacy. Instead of first computing \(y_i = aggr(\mathbf{r}_i)\), EM directly samples a private label \(\tilde{y}_i\) from the candidate response space \(S_i\). This makes it particularly suitable for questionnaire-based preference data, where the label is derived from a group distribution rather than directly provided as a single annotation.

\paragraph{DP-SGD.~\cite{Abadi2016DeepLW}}
We also evaluate DP-SGD where the label construction process stay unchanged and privacy protection is applied during fine-tuning. Specifically, per-example gradients are clipped to a fixed norm bound, and Gaussian noise is added to the aggregated gradient for each mini-batch. DP-SGD represents a standard approach for providing record-level privacy protection during model fine-tuning. However, in our setting, this can be overly conservative and introduces additional computational overhead.

Overall, the three mechanisms correspond to different privacy injection points in the alignment pipeline. RR perturbs the aggregated label before training, EM privatizes and replaces the aggregation step, and DP-SGD perturbs gradients during fine-tuning. Under the same WVS-based rating prediction task, this unified formulation allows us to compare their privacy-utility trade-offs and reveal the performance gap among different mechanisms applied at different stages, providing a privacy-efficient approach to LM preference alignment on questionnaire-style data.

\section{Experiments}\label{sec:experiment}
\subsection{Experimental Setup}
We evaluate the proposed privacy-preserving alignment pipeline on WVS-derived rating prediction tasks.
The dataset is constructed from 94 WVS seed questions \(q\), covering topics including economic development, democratization, religion, gender equality, social capital, and subjective well-being. Each seed question is rewritten 20 times with GLM-5~\cite{glm5team2026glm5vibecodingagentic}, and divided into a training set of 1,500 samples and a test set of 380 samples. All augmented samples derived from the same seed question are assigned to the same split to avoid data leakage.

Four settings, including non-private training, DP-SGD, RR, and EM, are evaluated on two Models: Llama-3.1-8B-Instruct~\cite{grattafiori2024llama} and Qwen3.5-9B~\cite{qwen3.5}. The model output is restricted to the discrete response space through a classification head for rating prediction task. Specifically, the non-private baseline uses deterministic Median aggregation on raw WVS responses \(y=Med(r_j)\).
We tune the learning rate, batch size, number of epochs, and LoRA rank when parameter-efficient fine-tuning is used. Batch size \(=32\), maximum epoch \(=5\), and LoRA rank \(=32\) are shared across all settings. A learning rate of \(1\times 10^{-5}\) is applied to RR and EM, and larger learning rate of \(5\times 10^{-4}\) is applied to DP-SGD, since gradient clipping and perturbation would otherwise suppress the effective update. Ablation study for hyperparameters and additional models are attached to Appendix~\ref{app:llamaresult} and~\ref{app:hyperparasel}
\subsection{Evaluation Metrics}
We evaluate each model in terms of prediction utility and privacy cost. Since WVS labels are ordinal ratings, we use normalized mean absolute error (NMAE) as the main metric:
\[
\mathrm{NMAE}=\frac{1}{N}\sum_{i=1}^{N}\left|y_i' - \hat{y}_i'\right|.
\]
Here, \(y_i'\) and \(\hat{y}_i'\) are the normalized values of the target and predicted labels, defined as \(y_i'=\frac{y_i-s_{\min}^{(i)}}{s_{\max}^{(i)}-s_{\min}^{(i)}}\) and \(\hat{y}_i'=\frac{\hat{y}_i-s_{\min}^{(i)}}{s_{\max}^{(i)}-s_{\min}^{(i)}}\), respectively. \([s_{\min}^{(i)}, s_{\max}^{(i)}]\) is the valid response range of the corresponding question. Lower NMAE indicates better rating prediction.

\subsection{Privacy Accounting}
We report \(\epsilon\) for all privacy-preserving methods as an indicator of private protection. For DP-SGD, \(\epsilon\) is computed using R\'enyi differential privacy~\cite{Mironov2017RnyiDP} accounting under \(\delta=5\times10^{-5}\). For RR and EM, \(\epsilon\) is directly given as a hyperparameter. Notably, the privacy budget depends on function sensitivity, the utility function, and the accounting method; therefore, identical \(\epsilon\) values are theoretical upper bonds and do not necessarily imply the same level of privacy protection across different mechanisms.

\subsection{Results}
\begin{figure*}[htb]
    \centering
        \includegraphics[width=0.6\linewidth]{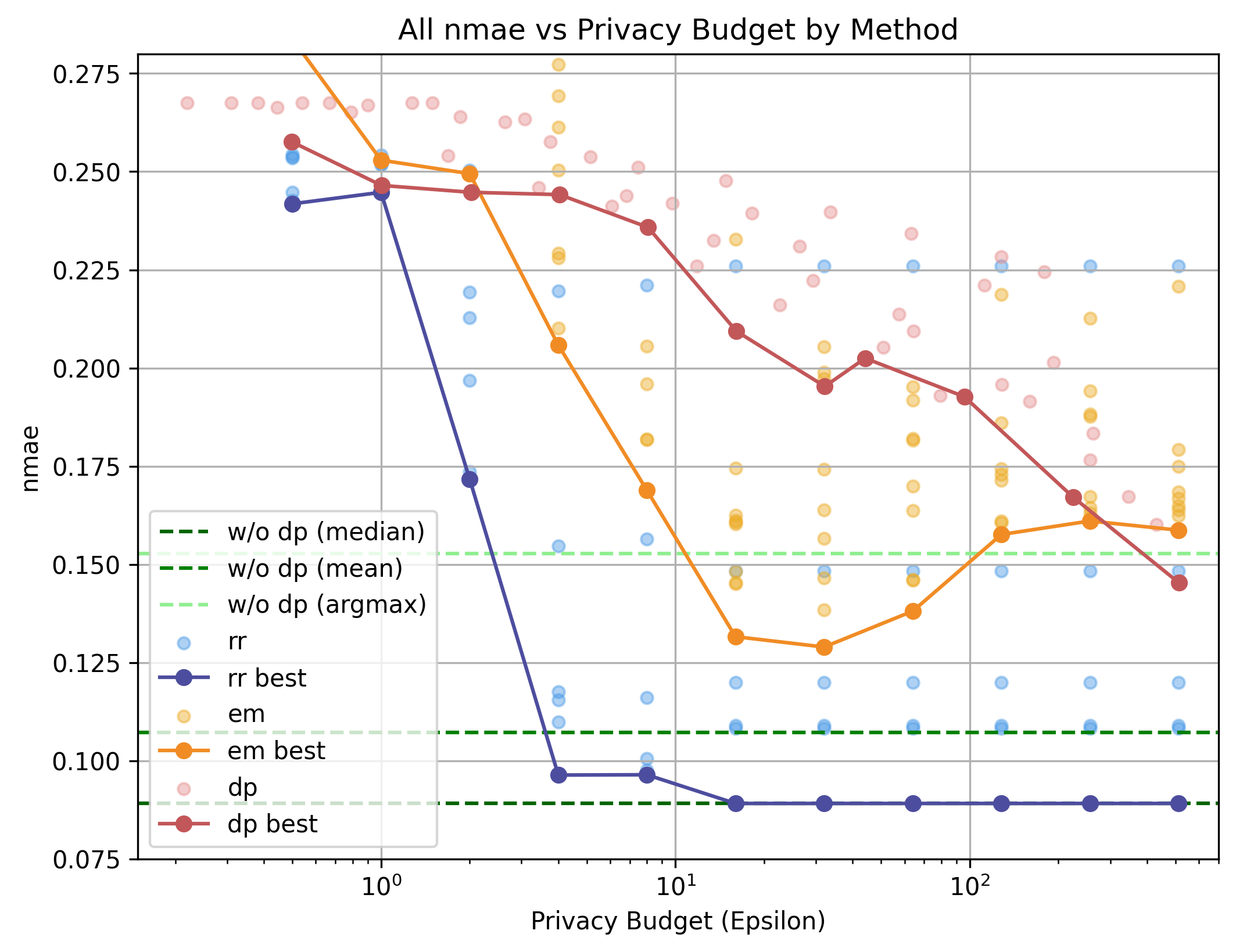}
    \caption{Best NMAE under different privacy budgets. Dashed horizontal lines denote non-private baselines with different aggregation functions; light scatters present suboptimal results.}
    \label{fig:privacy_utility_curves}
\end{figure*}

Fig.~\ref{fig:privacy_utility_curves} shows the privacy-utility trade-off of DP-SGD, RR, and EM under different privacy budgets on two backbone models. Each point in the curve reports the best validation performance, lighter scatter are sub-optimal results from other epochs, and dashed horizontal lines denote non-private baselines.
Overall, RR achieves the lowest NMAE across most privacy budgets. Its performance improves rapidly when \(\epsilon < 10\), and then converges to the non-private median baseline. This trend is expected under randomized response. When the response space size is \(k=4\), the probability of perturbing a label is approximately
\(p_{\mathrm{flip}} = \frac{3}{e^{\epsilon}+3}.\)
At \(\epsilon=9\), this probability is already smaller than \(1/2000\). Since the training set contains 1,500 samples, the expected number of flipped labels is less than one. Therefore, with larger \(\epsilon\), RR becomes nearly equivalent to non-private training.
Utility of DP-SGD continues to improve slowly as \(\epsilon\) increases. This indicates that gradient-level perturbation can preserve useful alignment signals when the privacy budget becomes larger, but compared with RR and EM, DP-SGD usually requires a much larger privacy budget to approach the optimal performance.

EM also improves rapidly at small privacy budgets and reaches its best performance at around \(\epsilon \approx 10\). However,  NMAE approaches argmax baseline because argmax can be viewed as a special variant of EM as \(\epsilon\) approaches infinity. This suggests that although EM can effectively privatize the aggregation step, its sampling process may introduce additional bias into the constructed labels. As a result, its best performance is slightly worse than RR, but it still provides a competitive privacy-preserving alternative to gradient-level DP, especially with specific privacy budget range.

Table~\ref{tab:best_nmae} summarizes the best NMAE achieved by each method on Qwen3.5 backbone model. We report the best NMAE, the corresponding privacy budget, and the gap to the non-private median baseline.
\begin{table*}[t]
    \centering
    \begin{tabular}{c|c|c|c|c}
        \hline
        Model & Method & Best NMAE \(\downarrow\) & Corresponding \(\epsilon\) & Gap to w/o DP \(\downarrow\) \\
        \hline
        \multirow{4}{*}{Qwen3.5-9B}
        & w/o DP & 0.0917 & N/A & 0.0\% \\
        & DP-SGD & 0.1453& 512.80& 63.08\%\\
        & RR & \textbf{0.0891}& 16.00& \textbf{0.00\%}\\
        & EM & 0.1289& 32.00& 44.71\%\\
        \hline
    \end{tabular}
    \caption{
    Best NMAE achieved by each method. The gap is computed relative to the non-private median baseline of the same backbone model.
    }
    \label{tab:best_nmae}
\end{table*}

\subsection{Aggregation Function Comparison}
We study the effect of 3 'M' aggregation functions for constructing ground-truth labels, they are: Median selected in main experiment (Sec~\ref{sec:experiment}) , compared with Mean and argMax to evaluate how the choice of aggregation function affects privacy-preserving alignment performance. Our key observation is that median + RR provides best performance among all combinations.

Mean and median have different statistical and privacy-related properties. Median aggregation is considered to be statistical robust, outperforming mean on outliers  and heavy-tailed response distributions. From the perspective of global sensitivity, however, mean has a theoretical advantage. In our setting, where labels are bounded in a discrete response space \(S\) and each question has \(n\) responses, the global sensitivity of mean aggregation is \(O(|S|/n)\). In contrast, the worst-case global sensitivity of median aggregation can be \(O(|S|)\), since changing a single response may shift the median substantially in adversarially constructed datasets. The derivations of these two sensitivity bounds are provided in Appendix~\ref{app:sensitivity}.

\begin{figure}
\centering

\parbox[b]{0.32\textwidth}{
\centering
\includegraphics[width=\linewidth]{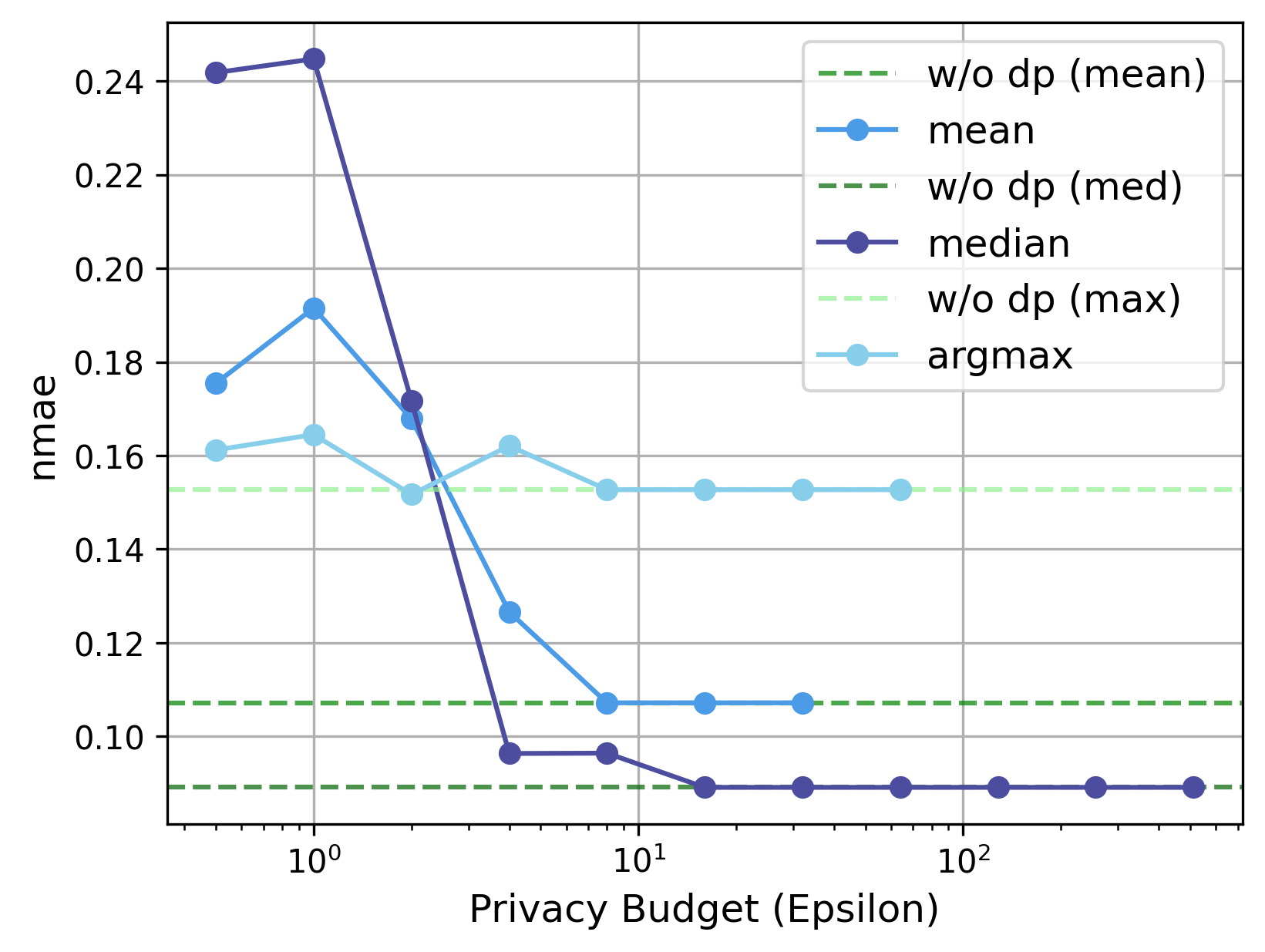}
\small (a) Random Response
}
\hfill
\parbox[b]{0.32\textwidth}{
\centering
\includegraphics[width=\linewidth]{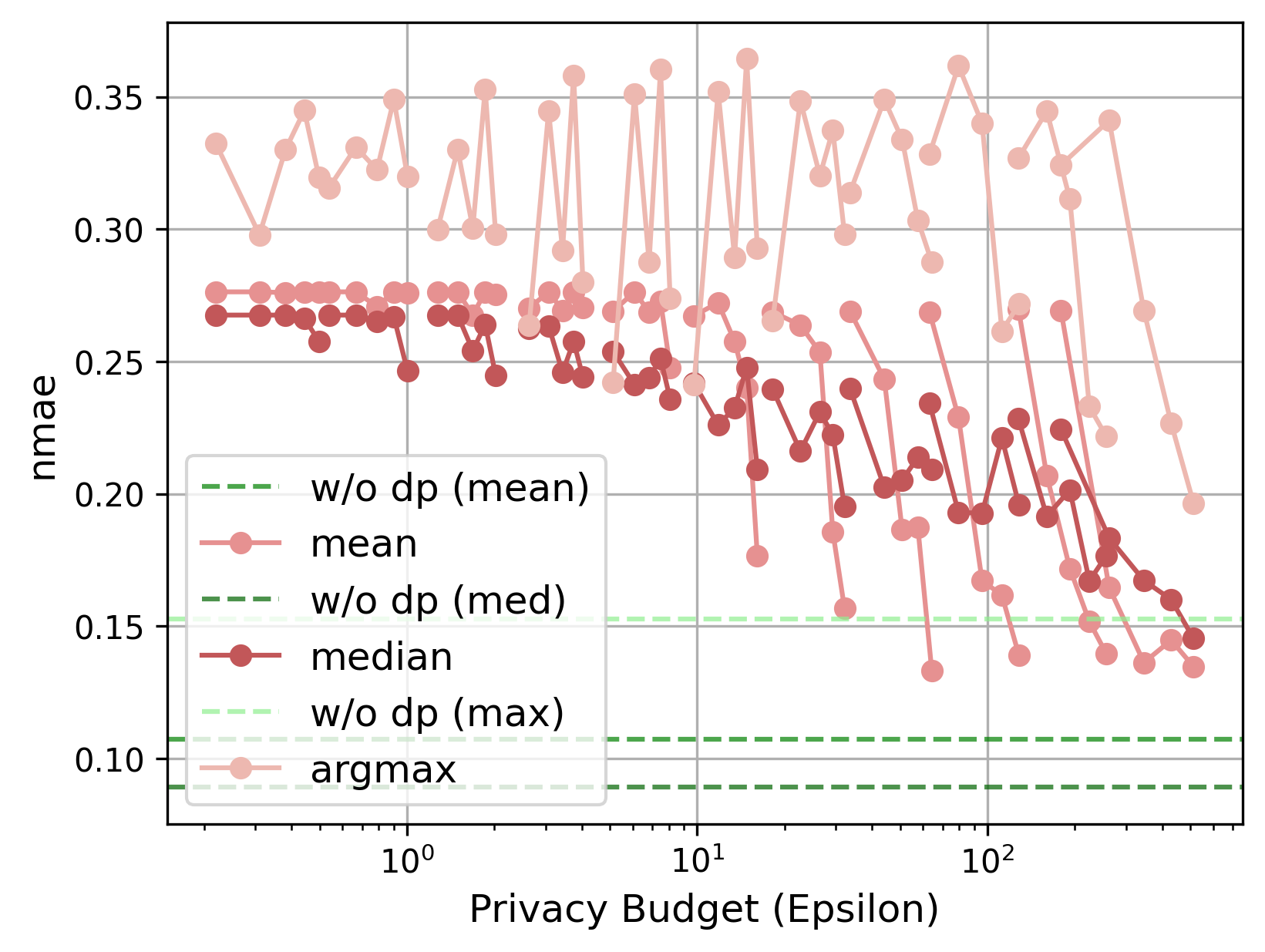}
\small (b) DP-SGD
}
\hfill
\parbox[b]{0.32\textwidth}{
\centering
\includegraphics[width=\linewidth]{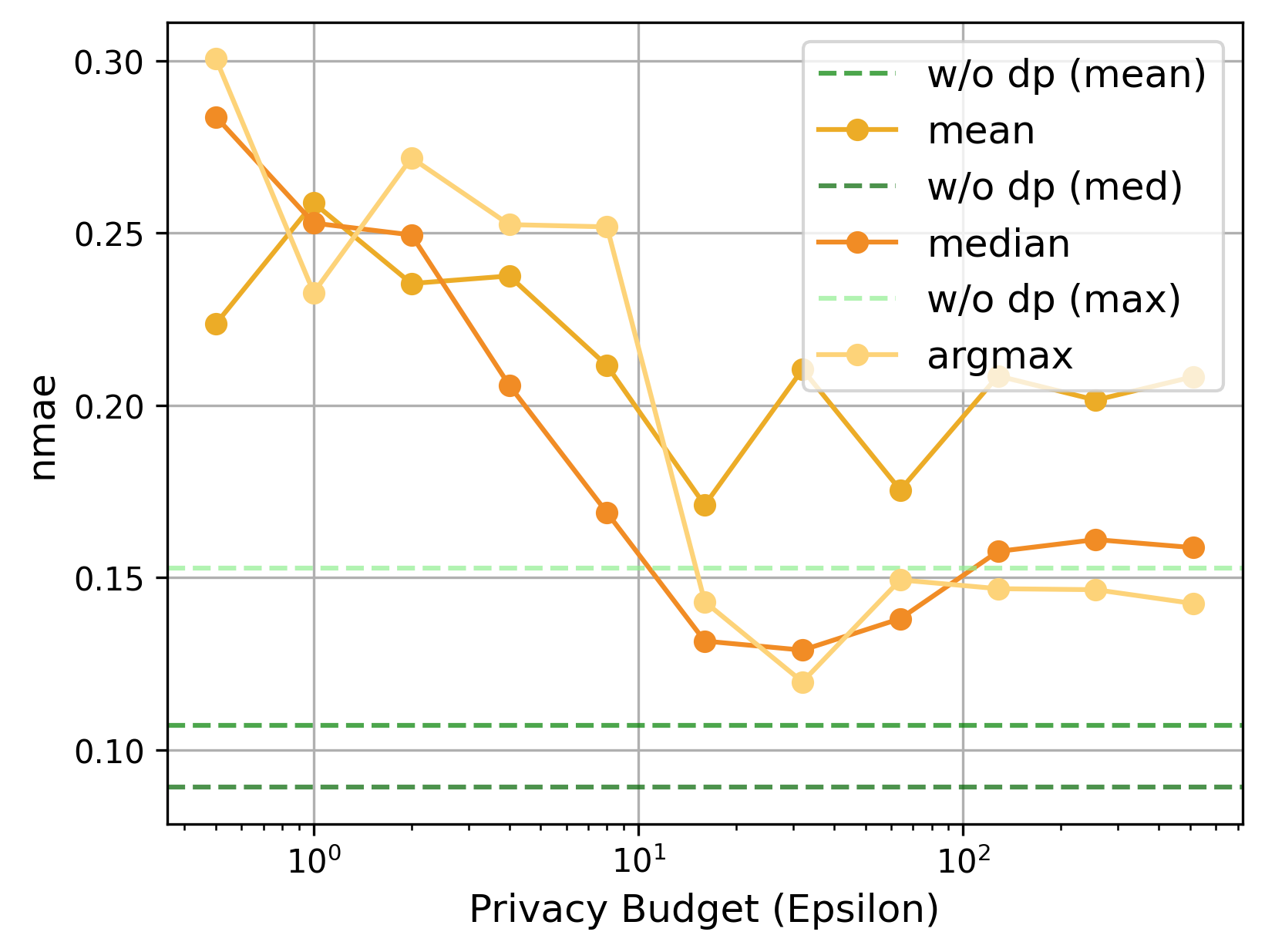}
\small (c) Exponential Mechanism
}

\caption{Privacy-utility comparison between mean, median and argmax aggregation.}
\label{fig:aggregate_function_comparison}

\end{figure}

Despite this theoretical difference, our empirical results show that median aggregation performs consistently better than or comparable to mean aggregation. As shown in Fig.~\ref{fig:aggregate_function_comparison}, under the same privacy budget, models trained with median-based labels achieve lower optimal NMAE than those trained with mean-based labels. This suggests that, although median has higher worst-case sensitivity, it may produce more stable and representative supervision signals for WVS-style ordinal responses. 
We also evaluate argmax aggregation, which selects the most frequent response label as the ground truth. As shown in Fig.~\ref{fig:aggregate_function_comparison}, argmax performs substantially worse than both mean and median. This may be because argmax discards the ordinal structure of the response space and only preserves the most frequent category, making it less informative when the response distribution is dispersed. Therefore, we use median aggregation as the default label construction method in the main experiments.

\subsection{Generalization: Generative Alignment on GOQA}
We also evaluate the generalization of the pipeline on GlobalOpinionQA (GOQA)~\cite{}, which provides per-country human opinion distributions over multiple-choice questions. Following the protocol of~\cite{}, we fine-tune the model with \(N\) sampled answers for each question, and evaluate the aligned model with next-token logits over option-letter tokens. Evaluation metrics are:  \(1-\text{JS}\) similarity (the primary metric of GOQA) and total variation (TV) between the model and human distribution. 

Results are reported in Table~\ref{tab:GOQAresult}. First, all label-level mechanisms show substantial improvement over DP-SGD and base model. Second, with larger privacy budget, RR (\(\epsilon>5\)) and EM (\(\epsilon>10\)) matches the non-private SFT baseline as expected. Additionally, model trained under large budgets learn over-concentrated distributions (average max-probability mass of 0.75, versus a target  model mass of 0.58), whereas strong DP constraints better preserve response  diversity (avg. mass \(< 0.51\)).

\begin{table}[t]
    \centering
    \small
    \setlength{\tabcolsep}{2.5pt}
    \begin{tabular}{c|cc|cccc|cccc|cc}
        \hline
         & Base & w/o DP & \multicolumn{4}{c|}{RR (\(\epsilon\)=1/5/10/20)} & \multicolumn{4}{c|}{EM (\(\epsilon\)=1/5/10/20)} & \multicolumn{2}{c}{DP-SGD} \\
        \hline
        \(1-\mathrm{JS}\) & 0.686 & 0.843 & 0.748 & 0.836 & \textbf{0.846} & 0.843 & 0.729 & 0.798 & 0.826 & 0.744 & 0.698 & 0.705 \\
        TV & 0.372 & 0.182 & 0.293 & 0.189 & \textbf{0.178} & 0.182 & 0.315 & 0.228 & 0.193 & 0.287 & 0.336 & 0.337 \\
        \hline
    \end{tabular}
    \caption{Alignment similarity between private aligned model and human distribution on GOQA. Best performance is marked as \textbf{bold}} 
    \label{tab:GOQAresult}
\end{table}

\subsection{Discussion}

Our results show that label-level DP mechanisms outperform classical DP-SGD, achieving lower best NMAE with smaller privacy cost. This is partly because label-level methods perturb only the sensitive preference labels, thereby avoiding redundant noise on the full training record. In addition, RR and EM provide direct and tight control over the privacy parameter, whereas the \(\epsilon\) of DP-SGD is estimated by an RDP accountant and serves as an upper bound on the true privacy cost.

Although EM exploits the ordinal structure of preference labels, it performs worse than RR in our experiments. One possible reason is the mismatch between its optimization target and the downstream learning objective: our EM formulation behaves  a randomized argmax, which may not be the most suitable aggregation rule for this task. This suggests that the design of the EM utility function, analogous to loss-function design, is critical for improving utility.

Finally, our results highlight the importance of aggregation function design. Median aggregation shows stronger empirical robustness than mean and argmax, suggesting that it better extracts stable preference signals from noisy response distributions. Future works of directly utilizing raw distribution instead of aggregated labels are of great importance.

\section{Conclusion}
We study differential privacy for preference alignment with questionnaire-based group preferences. Our main contribution is a simple privacy-preserving pipeline that adapts standard alignment procedures to this setting while providing formal privacy guarantees for the protected preference labels. The pipeline is compatible with data-specific protection mechanisms. Experimental results on WVS and GOQA show that lightweight label-level privacy mechanisms provide tighter privacy control, stronger alignment utility, and better training efficiency than the record-level baseline. These findings suggest that the proposed pipeline is a promising approach to privacy-preserving alignment when supervision signals are derived from socially sensitive survey data.
  
\section{Limitations}This study has several limitations. First, our evaluation primarily focuses on privacy-utility trade-offs over preference prediction tasks, while aligned models behaviors on broader downstream tasks remain underexplored. Second, the privacy budget \(\epsilon\) is instantiated differently in three DP mechanisms, and the upper bounds reported provide only limited practical protection at large values. Third, although EM leverages the ordinal structure of labels, it is relatively simple and may not fully capture the structure of survey-derived preferences. Finally, Our motivating attack operates at the group level rather than identifying individual records; extending the attack analysis and protection to respondent-level granularity remains future work.


\bibliography{ref}
\appendix
\section{Llama Result}\label{app:llamaresult}
This section shows additional result on Meta-Llama-3.1-8B-Instruct for Section~\ref{sec:experiment}.

\begin{figure}[htb]
\centering

\parbox[b]{0.32\textwidth}{
\centering
\includegraphics[width=\linewidth]{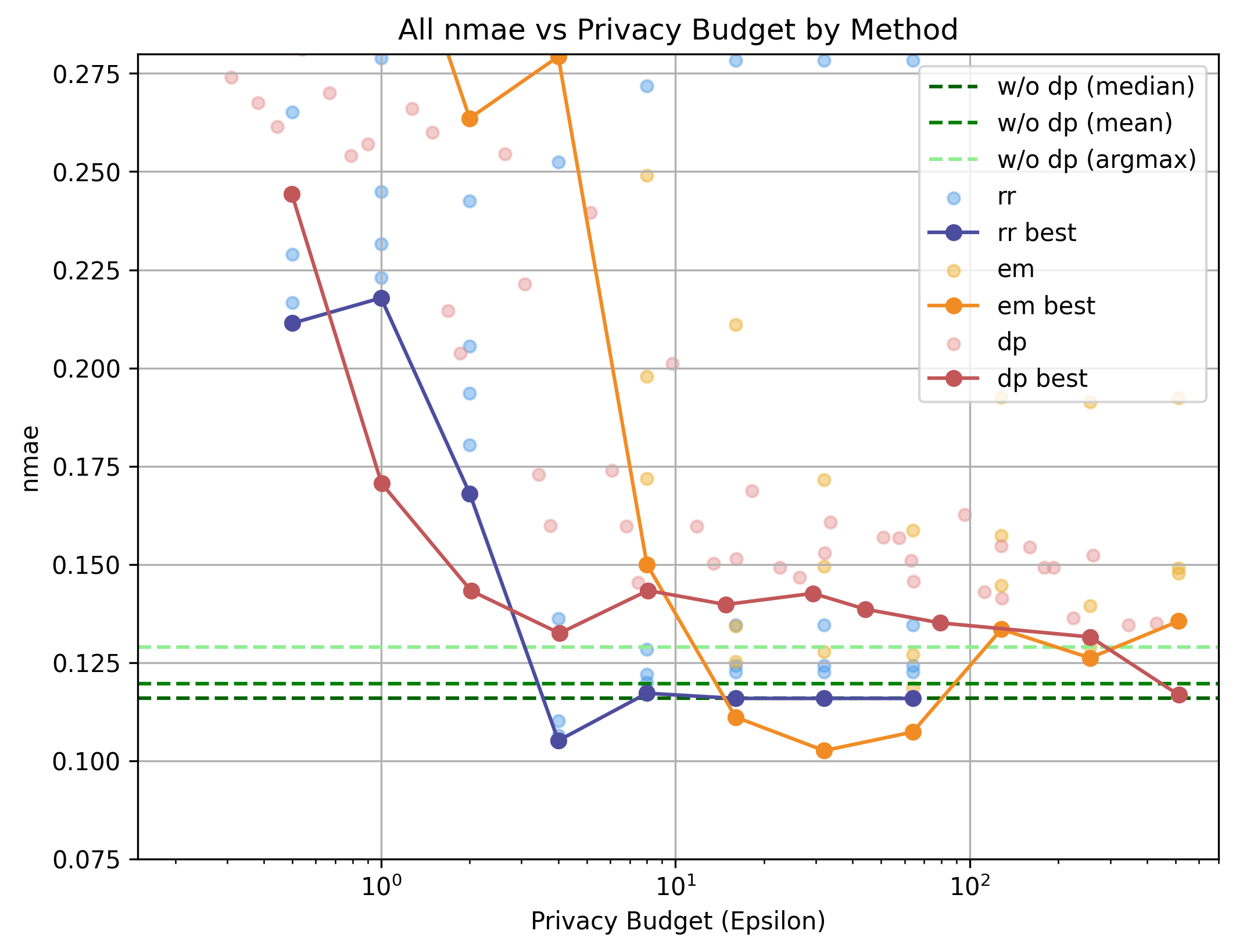}
\small (a) Median
}
\hfill
\parbox[b]{0.32\textwidth}{
\centering
\includegraphics[width=\linewidth]{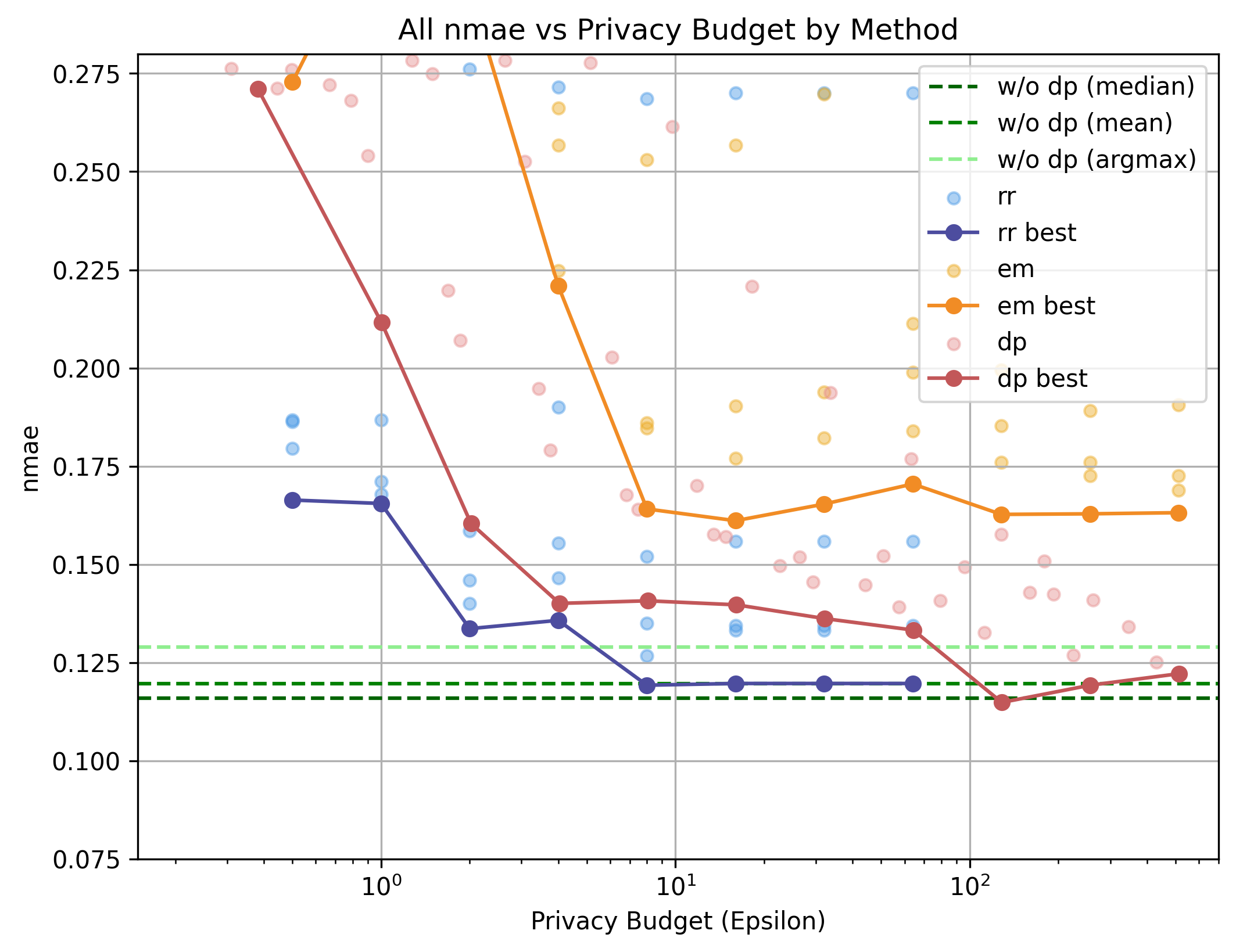}
\small (b) Mean
}
\hfill
\parbox[b]{0.32\textwidth}{
\centering
\includegraphics[width=\linewidth]{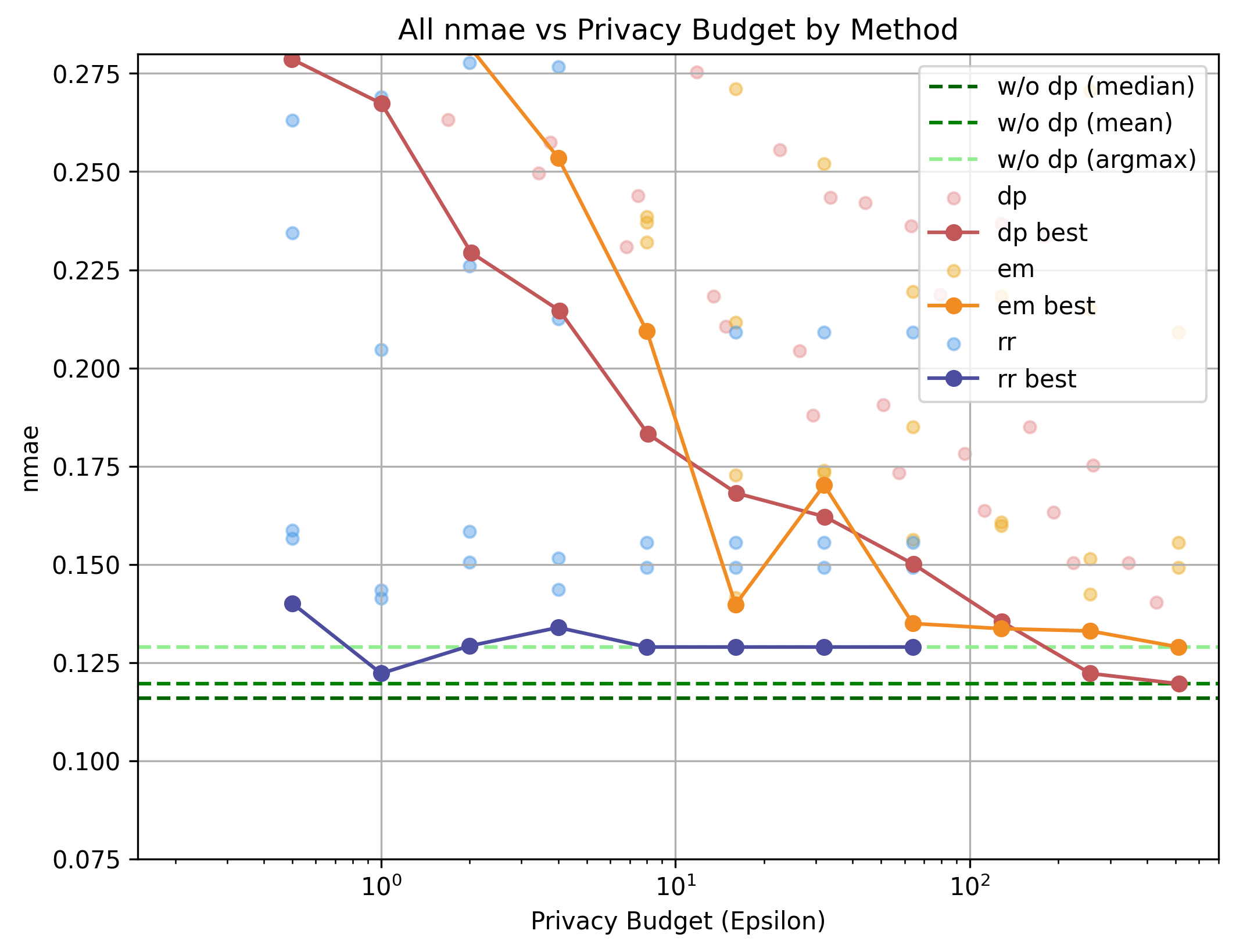}
\small (c) Argmax
}

\caption{Privacy-utility plot on Llama-3.1-8B-Instruct with three aggregate functions}
\label{fig:llama_privacy_utility}

\end{figure}

Specifically, in Fig.~\ref{fig:llama_privacy_utility}, EM and RR achieve their best performance below the no DP baseline, which may seem counterintuitive since baseline add no perturbation. Two explanations are possible: median aggregation only provides a thin aspect of the raw preference data, and proper noise added by RR or EM may allow the model to learn the hidden distribution and introduce generalization ability; second, single-run experiment with fixed random seed could also contribute to the bias.
 
\section{Selection of Hyperparameters}\label{app:hyperparasel}
\begin{figure}[htb]
\centering

\parbox[b]{0.24\textwidth}{
\centering
\includegraphics[width=\linewidth]{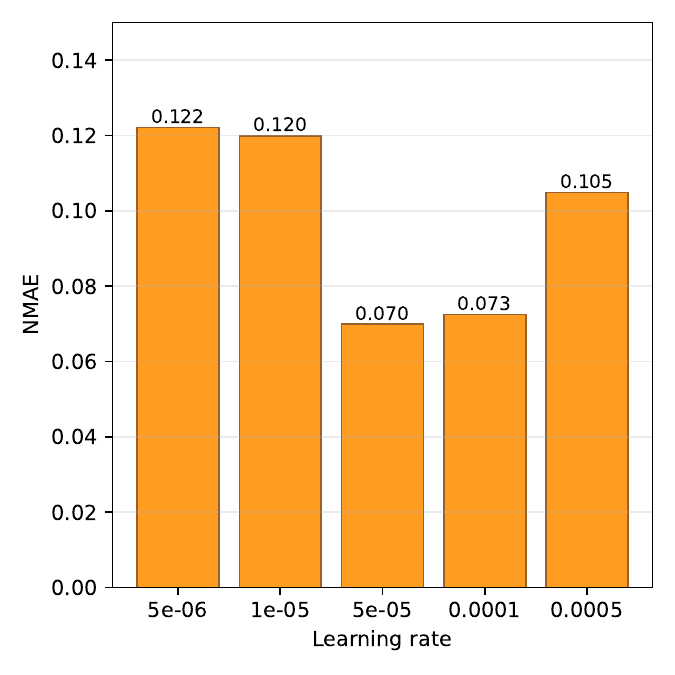}
\small (a) Learning rate
}
\hfill
\parbox[b]{0.24\textwidth}{
\centering
\includegraphics[width=\linewidth]{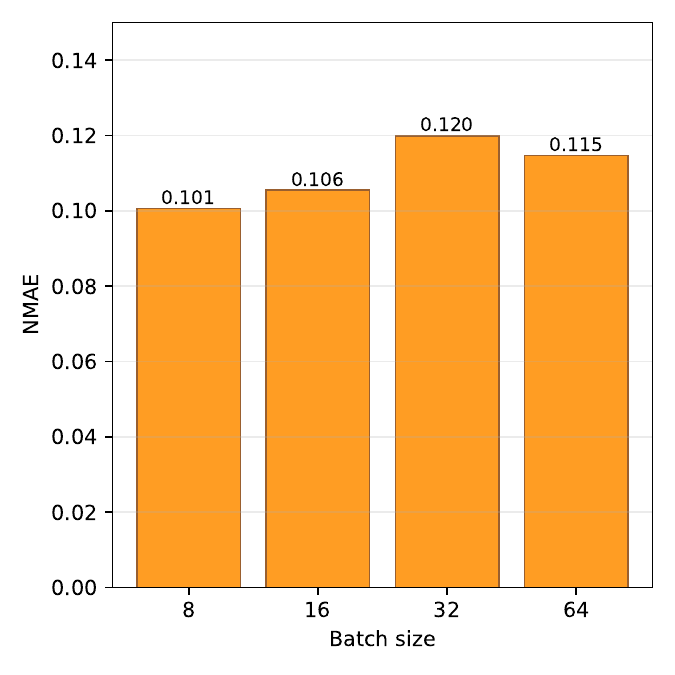}
\small (b) Batch size
}
\hfill
\parbox[b]{0.24\textwidth}{
\centering
\includegraphics[width=\linewidth]{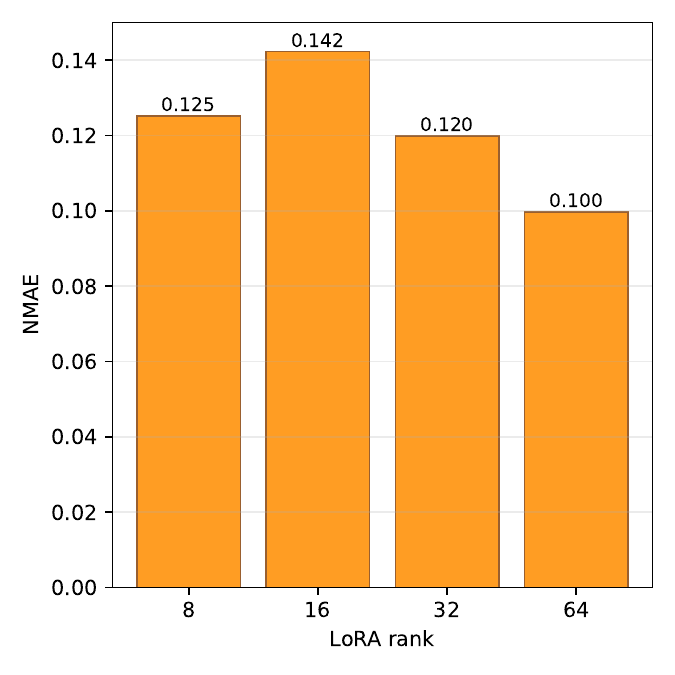}
\small (c) LoRA rank
}
\hfill
\parbox[b]{0.24\textwidth}{
\centering
\includegraphics[width=\linewidth]{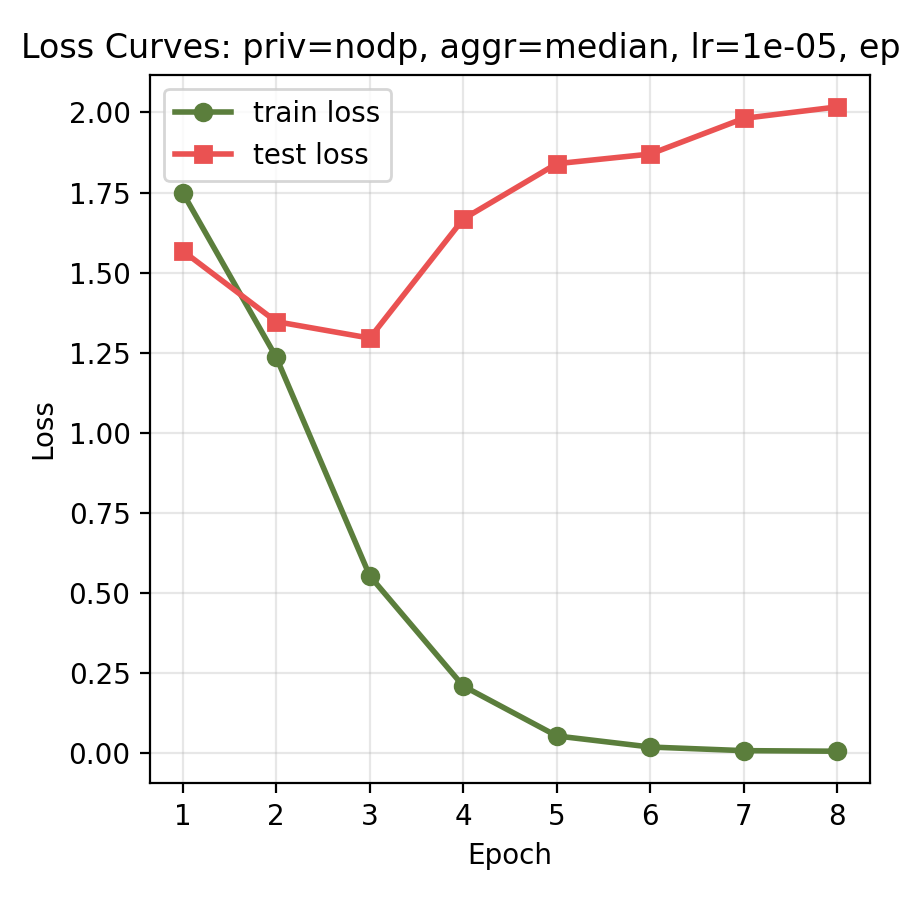}
\small (d) Max epoch
}
\caption{Hyperparameter ablation results. Panels (a)-(c) show the best NMAE across epochs for different hyperparameter values, while panel (d) reports the train/test loss curves over epochs.} 
\label{fig:hyperparameter_ablation}
\end{figure}

We determine the learning rate, batch size, number of training epochs, and LoRA rank through ablation studies. As shown in Fig.~\ref{fig:hyperparameter_ablation}, we select the configuration that provides the best trade-off between validation utility, training stability, and computational cost. The resulting hyperparameters are fixed for all subsequent experiments unless otherwise stated.

\section{Sensitivity of Aggregation Functions}\label{app:sensitivity}
Sensitivity determines how much the output of a function can change when a single input record is modified, and thus controls the scale of noise required by many DP mechanisms.

Let $D=(r_1,\ldots,r_n)$ be a response vector with $r_i\in[L,U]$. Two datasets $D$ and $D'$ are neighboring if they differ in exactly one response \(r_j\ne r_j'\). The global sensitivity of an aggregation function $u$ is
\[
\Delta u
=
\max_{D\sim D'}
|u(D)-u(D')|
\]
\paragraph{Mean.}
For \(\mathrm{mean}(D)=\frac1n\sum_{i=1}^{n} r_i\), changing one response modifies the output by at most
\[
\left|
\mathrm{mean}(D)-\mathrm{mean}(D')
\right|
=
\frac{|r_j-r_j'|}{n}
\le
\frac{U-L}{n}
\]
The bound is tight by setting $(r_j,r_j')=(U,L)$, hence
\[
\boxed{
\Delta_{\mathrm{mean}}
=
\frac{U-L}{n}
}
\]
\paragraph{Median.}
For odd $n=2m+1$, consider
\[
D=(\underbrace{L,\ldots,L}_{m},
\underbrace{U,\ldots,U}_{m+1})\\
D'=(\underbrace{L,\ldots,L}_{m+1},
\underbrace{U,\ldots,U}_{m})
\]
which differ in one entry. Then
\[
\mathrm{med}(D)=U,\qquad
\mathrm{med}(D')=L
\]
yielding
\[
\boxed{
\Delta_{\mathrm{median}}
=
U-L
}
\]
For even $n=2m$, using the standard definition \(\mathrm{med}(D)=\frac{x_{(m)}+x_{(m+1)}}{2}\), a neighboring construction gives
\[
\boxed{
\Delta_{\mathrm{median}}
=
\frac{U-L}{2}
}
\]
\paragraph{Implication.}
Therefore,
\[
\Delta_{\mathrm{mean}}
=
O\!\left(\frac{U-L}{n}\right),
\qquad
\Delta_{\mathrm{median}}
=
O(U-L)
\]
Under global-sensitivity-based DP mechanisms, mean aggregation requires substantially less noise than median aggregation. Nevertheless, as shown in Section~5.5, median may still achieve better empirical utility because global sensitivity alone does not characterize label quality or robustness.

\end{document}